\documentclass[sigconf]{acmart}

\usepackage{subfig}
\usepackage{balance}

\AtBeginDocument{%
  \providecommand\BibTeX{{%
    \normalfont B\kern-0.5em{\scshape i\kern-0.25em b}\kern-0.8em\TeX}}}

\setcopyright{acmcopyright}
\copyrightyear{2020}
\acmYear{2020}

\begin{document}

\title{Pose Estimation for Facilitating Movement Learning from Online Videos}

\author{Atima Tharatipyakul}
\affiliation{%
  \institution{Singapore University of Technology and Design}
}
\email{atima\_tharatipyakul@mymail.sutd.edu.sg}

\author{Kenny Choo}
\affiliation{%
  \institution{Singapore University of Technology and Design}
}
\email{kenny\_choo@sutd.edu.sg}

\author{Simon T. Perrault}
\affiliation{%
  \institution{Singapore University of Technology and Design}
}
\email{perrault.simon@gmail.com}

\renewcommand{\shortauthors}{Tharatipyakul, et al.}

\begin{abstract}
There exist a multitude of online video tutorials to teach physical movements such as exercises. Yet, users lack support to verify the accuracy of their movements when following such videos and have to rely on their own perception. To address this, we developed a web-based application that performs human pose estimation using both video inputs from the online video and web camera, then provides different types of visual feedback to a user. Our study suggests that the user's skeleton overlaid on the user's camera feed improved user performance, whereas the user's skeleton on its own or trainer's skeleton with the trainer video offered limited benefits. We believe that our application demonstrates the potential to enhance learning physical movements from online videos and provides a basis for other guidance systems to design suitable visualizations.

\end{abstract}



\begin{teaserfigure}
  \includegraphics[width=\textwidth]{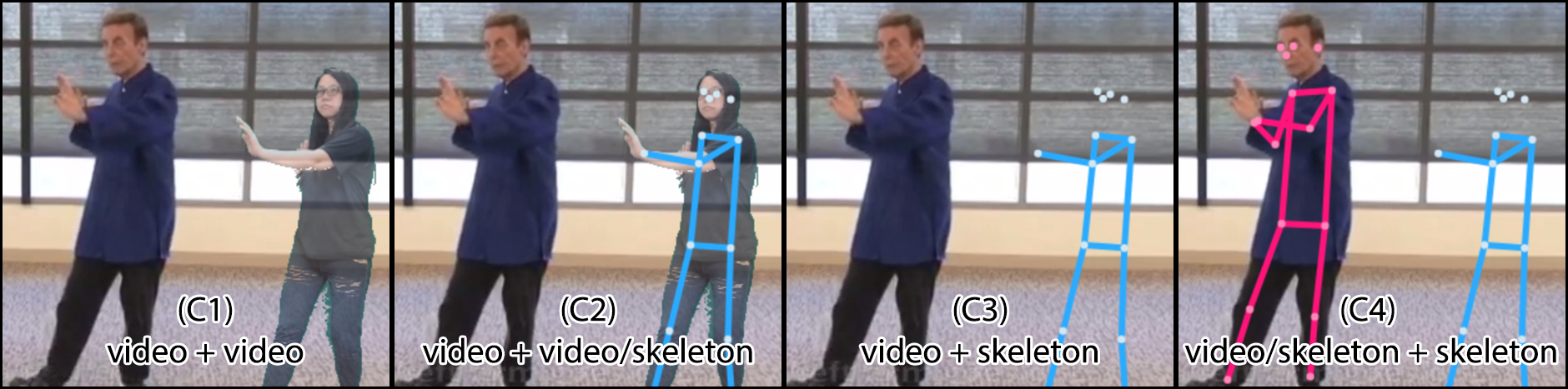}
  \caption{A study on four different types of visual feedback on a tai chi video: (C1) trainer video + user video, (C2) trainer video + user video with skeleton, (C3) trainer video + user skeleton, and (C4) trainer video with skeleton + user skeleton. The tutorial video is from https://youtu.be/ZxcNBejxlzs.}
  \label{fig:teaser}
\end{teaserfigure}

\maketitle

\section{Introduction}

Online video tutorials have become a popular mean to learn physical movement. The production of the online videos, however, could range from professional editing, which considers aspects such as viewer perspectives and additional graphics, to the simplest from-the-phone recordings. The unconstrained nature of online videos makes it hard for automated processing like extracting a trainer's pose. Physical movement learning tools that require the trainer's pose to provide guidance, such as those in \cite{Sigrist2013, 01b481b8931b49faa196321809e22d89, viglialoro2019review}, rarely use online videos as a basis for training even though they are massively available and untapped resource.

The augmentation of online videos with useful feedback presents a promising mean to improve physical movement learning, as seen in previous works with the Microsoft Kinect (e.g.  \cite{anderson2013youmove,Lin:2018:SME:3265845.3265856}). Still, only limited number of works address different ways to represent the trainer and user (e.g. 3D model against skeleton \cite{8949226}). Advances in computer vision with unconstrained camera feeds (e.g. \cite{cao2018openpose,posenet}) motivate us to reexamine online videos and augment them with pose information extracted from the videos.
How different visualizations could facilitate movement learning from online videos remains as an open issue and is our main research question.

In this paper, we created a system that uses pose estimation to provide visual feedback to users who want to follow tutorial videos. The contributions of our study include (1) the development of a fully-working interactive system featuring configurable interface to facilitate learning from online videos and (2) the study and insights on four types of visual feedback, relying on combinations of (a) the video feeds of both video trainers and the users and (b) pose estimation data represented as a skeleton. We used tai chi, a physical activity that requires highly accurate movements, as a case study in our controlled experiment. Our results suggest that participants were able to follow tutorials more closely using the trainer video with their own video feed that superposed with pose estimation data (skeleton).

\section{Background and Related Work}

\emph{Tai Chi.} Tai chi is a physical activity with slow continuous movements to promote well-being. As it requires highly accurate movements, it has been popularly used in research to help users practice it. Due to the complexity of the movements, most works employ virtual 3D teachers, either with a screen \cite{de2011serious}, head mounted displays \cite{iwaanaguchi2015cyber, Han:2017:MTC:3041164.3041194, 8797986, chua2003training, hulsmann2018classification}, or audio-visual-tactile devices \cite{portillo2008real}. The most similar work to ours is Stillness Moves \cite{Lin:2018:SME:3265845.3265856}, which uses the recorded weight from pressure sensing shoes and video data from the Kinect for the training. However, input from the Kinect, webcam, and other camera-based devices tend to suffer from occlusion and/or estimation error. Therefore, trainer data are usually collected in controlled settings to ensure its accuracy, as seen in \cite{Lin:2018:SME:3265845.3265856} and other systems (e.g.\cite{iwaanaguchi2015cyber}). While there exist works on other type of physical activity that capture data in unconstrained scenes \cite{khurana2018gymcam}, none of them address edited videos, nor compare multiple types of visual feedback as we do in this paper.

\emph{Visual Feedback Design.} Sigrist et al. \cite{Sigrist2013} reviewed different types of feedback strategies such as concurrent feedback, terminal feedback, fading feedback, performance-based feedback, self-controlled feedback, and bandwidth feedback. To help the user verify and improve their movement, the learning tools must provide a mean to compare the user's data with the trainer's. Hence, we can also categorize the visualization based on comparison approaches \cite{Gleicher01102011}: juxtaposition (putting objects side-by-side), superposition (overlaying objects on top of each other), and explicit encoding (representing relationship directly). Each approach or strategy has its own strengths and weaknesses. For instance, Timmermans et al. suggested concurrent feedback to be effective for beginners while terminal feedback may benefit more skilled users \cite{timmermans2009technology}. We use these insights to justify our design choices.

\emph{Video Feed vs. Skeleton Information.} Despite the capabilities of the visual domain, reduced feedback visualization could prevent an overload of information for complex motor tasks \cite{eaves2011short}. One essential consideration when dealing with videos, which are considered as complex objects for comparison, is how to reduce the complexity of the videos to reduce cognitive load \cite{8017615,Tharatipyakul2018}. Displaying the skeleton or skeleton on top of video feed, as seen in \cite{anderson2013youmove,Lin:2018:SME:3265845.3265856} and many systems, could bring interested information to the user and reduce the cognitive load. The skeleton, however, may result in lose details and hider feedback effectiveness. For instance, Haoran et al.  \cite{8949226} found 3D models to be more effective than skeleton in supporting core learning. Still, none of the work mentioned above investigated the effectiveness of different visual feedback using video feed and/or skeleton information from pose estimation. Our work is thus the first study to perform such a comparison. 

\section{Feedback and System Design}




\subsection{Visual Feedback Design}
We wanted our system to provide high quality feedback that would allow users to follow a tutorial video accurately. The idea was to add relevant pieces of information on the user's current pose to enable a quick side-by-side comparison.

Showing the user's body is an easy and convenient way to provide accurate feedback. After some initial testing, where the user's video feed was located too far away from the trainer, participants reported that it was hard to follow the trainer as they had to constantly look at the trainer, then their own feed sequentially. We realized that both video feeds had to be shown as closely as possible, to keep all the relevant information in the user's central vision. We decided to use background subtraction. This way, we can reduce the distance between the feeds, and the user can focus on the important parts of the video feed, i.e. their own body.

In addition to the user's video feed, we also decided to use pose estimation data. Previous work~\cite{anderson2013youmove,Lin:2018:SME:3265845.3265856} made use of that data by drawing a skeleton to represent the user, showing  the individual position of the joints on the user's body (e.g. shoulder, legs, arms).

\subsection{Implementation}

\begin{figure}[hbt]
  \centering
  \includegraphics[width=\linewidth]{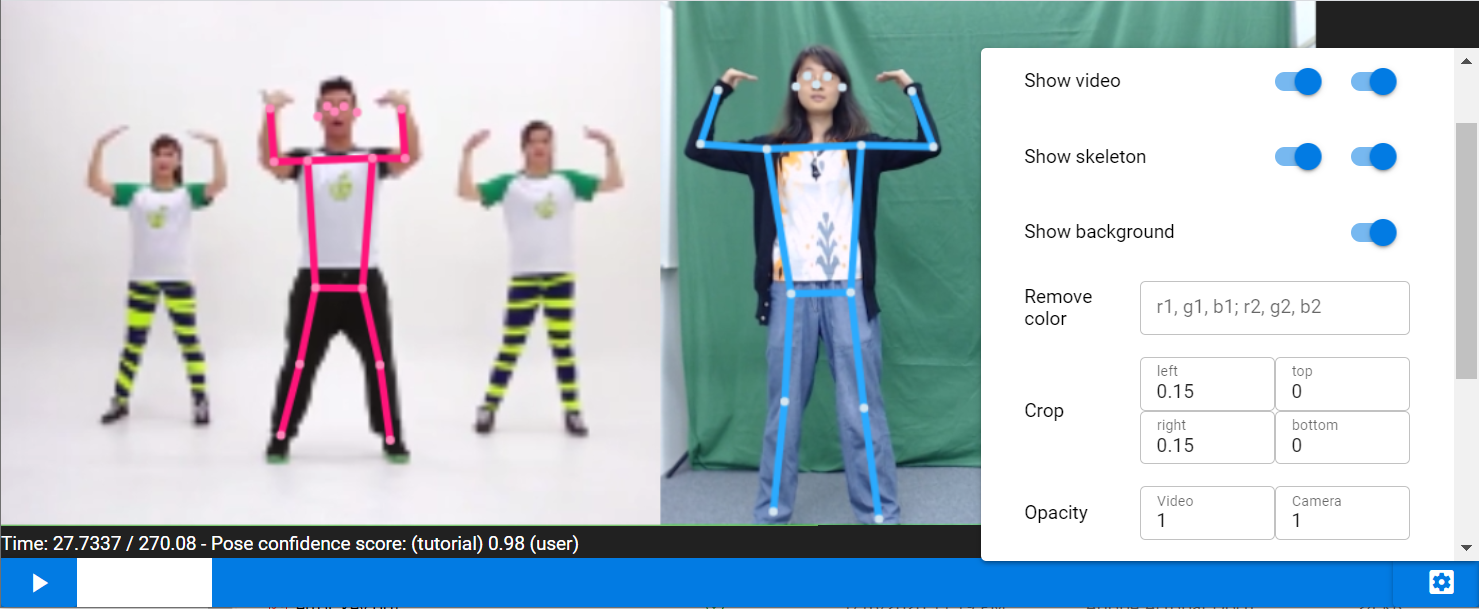}
  \caption{Screenshot of our application with options to control visibility and appearance of the videos and skeletons. The tutorial video is from https://youtu.be/3EbBHoRgn-Y.}
  \label{fig:ss}
\end{figure}

We developed a web-based application that takes the URL of a tutorial video file and the user's web camera feed as inputs.
The overall interface is similar to a typical video player, except that it also provides concurrent visual feedback -- showing a mirror image of the user from the camera beside the trainer in the video, as demonstrated in Figure~\ref{fig:ss}.

The application supports three types of content visualization: video, filtered video, and skeleton.
We support various functions such as cropping of the video frames, adjusting opacity, or removing the background (either using chroma keying with a green screen or automated person segmentation \cite{bodypix}).
The skeleton is estimated in real-time using a state-of-the-art pose estimation library~\cite{posenet}.
The application supports individual visibility toggling of the video feed and the skeletons of either the user or trainer.

The application facilitates comparison of the trainer and the user through natural feedback visualization. The user can compare themselves with the trainer side by side. Both feeds can be moved around, resized, and overlaid on top of another. In contrast with explicit encoding design that requires predefined pose comparison algorithm, these juxtaposition and superposition approach leverages user's perception to realize different between their movement and the trainer's.

\section{User Study}
We conducted a controlled experiment on each visualization type on a tai chi video to examine how each visualization type works.
Our goal here was not testing all possible combination of configurations, but finding out usefulness of a trainer's skeleton, user's skeleton, and user's video feed. The four conditions we selected for our user study are: (C1) trainer video + user video (baseline), (C2) trainer video + user video with skeleton, (C3) trainer video + user skeleton, and (C4) trainer video with skeleton + user skeleton, as illustrated in Figure~\ref{fig:teaser}.
By limiting number of conditions to four, we did not exhaust or overload participants with too many choices.

\subsection{Participants}
We used Latin Square to minimize the number of required participants (i.e. 4 participants for 4 conditions) and replicated it 3 times to increase the degrees of freedom for experimental error (see \cite{allen2017sage}). In total, we recruited 12 participants (7 female), aged from 22 to 37 years old ($M=27.50$, $SD=4.56$) for the experiment. Only one of them had rarely practiced tai chi.

\subsection{Apparatus}
For this experiment, we used a PC with a Xeon E5-2603 processor at 1.6 GHz, with 16 GB of RAM and a Nvidia Titan GPU with experimental software written in JavaScript running on a Chrome web browser. For the user video input, we used a Logitech C922 Pro webcam and scaled the resolution to 640 $\times$ 360 to match with a tutorial video retrieved from YouTube and maintain a good frame rate with the pose estimation. The computer was connected to a 27" monitor with a resolution of 2560 $\times$ 1440 pixels. We also used a green screen located behind the user to perform the background subtraction in real time.

\subsection{Procedure}
We first obtained informed consent from the participants and got them to fill in a demographics survey. They were then briefed about the experiment and continued to perform the four trials -- one for each visual feedback condition (C1, C2, C3, C4).
At the beginning of the trial, we asked participants to stand on a mark on the floor located 2.5 meters away from the monitor.
This was to ensure that the camera captured the whole body of the participant, and the videos were within a participant's central vision.
Participants were then allowed to try the condition for a period of 30 to 60 seconds for training, before proceeding with the actual trial.
We used the same video\footnote{https://youtu.be/zDSlx3INZWk} for the training for all conditions.

Tai chi tutorial video would then start, and the participant had to follow the movements as closely as possible. Each trial contained one move and lasted for about one minute. We chose four different moves of similar difficulty and presentation from one video\footnote{https://youtu.be/ZxcNBejxlzs}. The trainer repeated each move 6 times (3 times for left and right side). The original audio in the video was muted to focus participants on the visual feedback.
After finishing a trial, participants had to fill a NASA TLX questionnaire and provide subjective feedback about the condition.
After completing all four conditions, participants were asked to rank the techniques and indicated where they looked during the trials (experimenter, themselves, or both).



\subsection{Design}
We used a within-subject design with one independent variable: \emph{Visual Feedback} \{ (C1) trainer video + user video (baseline), (C2) trainer video + user video with skeleton, (C3) trainer video + user skeleton, and (C4) trainer video with skeleton + user skeleton \}. The order of presentation of the visual feedback condition was counterbalanced using a Latin Square.

We measured the angular error as a dependent variable. We computed the average angular difference of the shoulder, hip, upper and lower arms, as well as upper and lower legs between the trainer and the participant. The angular error was measured in radians and was not displayed to the participant. We excluded the head from the calculation since the participant had to tilt their head when checking the pose on-screen.
We also measured the individual components of the NASA TLX as dependent variables.

Participants completed the experiment in about 60 minutes and were encouraged to take breaks between conditions. Our design is as follows: 12 participants $\times$ 4 conditions $\times$ 1 repetition = 48 trials.

\subsection{Results}
\subsubsection{Error}
\begin{figure}[htb]
  \centering
  \includegraphics[width=\linewidth]{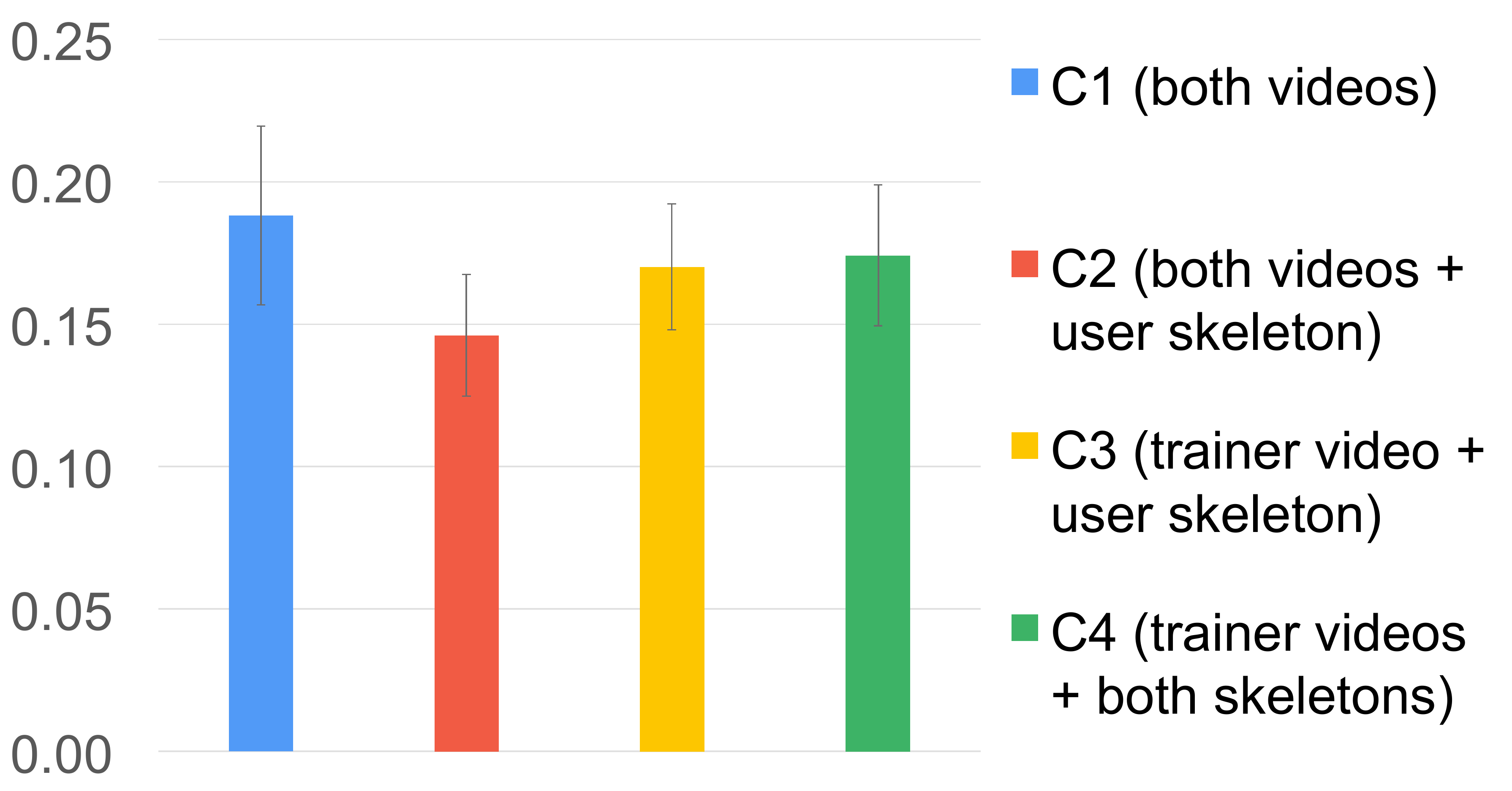}
  \caption{Average angular error in radians across conditions. Error bars show .95 confidence intervals.}
  \label{fig:error}
\end{figure}

We used one-way ANOVA for statistical analysis. We found a significant main effect of \emph{Visual Feedback} on error ($F_{3,33}=3.63$, $p=.02$). The average angular error was 0.169 radians. A Tukey HSD test for post-hoc analysis showed that the trainer video + user video with skeleton (C2) condition achieved the lowest error ($M=0.146~rad$), which was significantly lower than our baseline (C1, both videos, $M=0.188~rad$, $p<.05$). We did not find any other significant differences in terms of error, with our other conditions (C3 and C4, see Figure~\ref{fig:teaser}) achieving an accuracy of 0.170 and 0.174 rad respectively. The error rates are summarized in Figure~\ref{fig:error}.

\subsubsection{NASA TLX}
\begin{figure}[hbt]
  \centering
  \includegraphics[width=\linewidth]{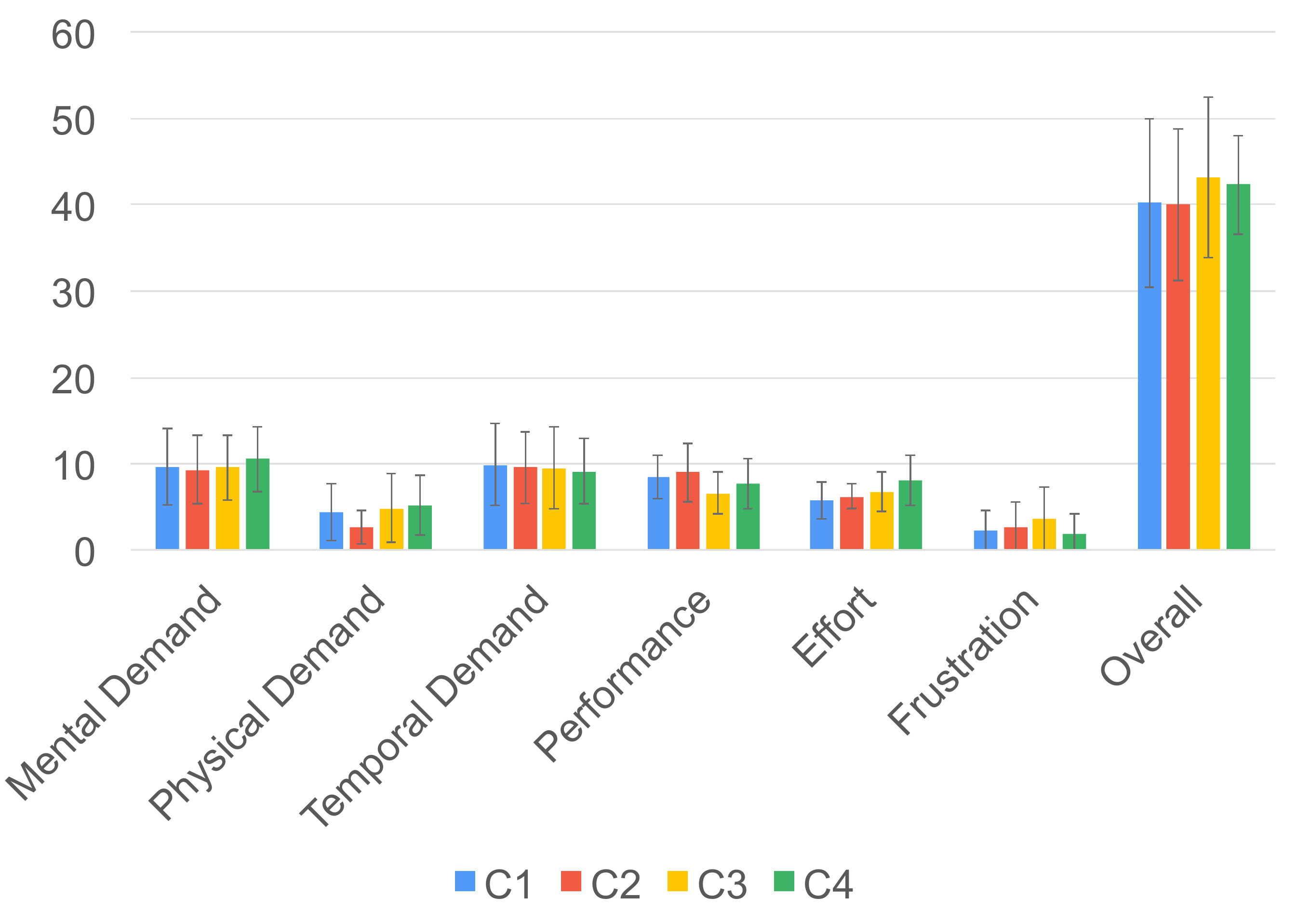}
  \caption{Average NASA-TLX for each condition.  Error bars show .95 confidence intervals.}
  \label{fig:tlx}
\end{figure}

Each condition achieved an average score between 33.8 (C2) and 35.9 (C1 and C4). We did not find any significant main effect of \emph{Visual Feedback} on the overall TLX score, nor with any individual component, except for Physical Demand ($F_{3,33}=3.4$, $p=.03$). Specifically, C2 ($M=2.56$) was as deemed as significantly less physically demanding as compared to C4 ($M=5.19$, $p<.05$). The NASA TLX scores are shown in Figure~\ref{fig:tlx}.

\subsubsection{Ranking}
In addition, C2 was ranked first or second best technique by 9 participants out of 12. We decided to sum the rank given by participants to each technique (1st = 1 point, 2nd = 2 points, etc.), with the lower the score, the better. By doing so, we found that C2 gets the lowest (best) score of 24, followed by C4 with 28 points, C1 with 33 points and C3 with 35 points.


\subsubsection{Subjective Feedback}
Participants liked C1 (baseline) and found it "easy" (P4) and "clear to follow" (P2, P12). Participants liked being able to compare their live video with the trainer's as it helped them "correct movements" (P5, P8, P10), but was deemed as "distracting" (P7).
Participants enjoyed the addition of the skeleton on themselves as it helped them focus on their own movements (P4).
Participants reported that C3, where the user's video feed is not displayed, made it harder to "compare" movements with the trainer (P1, P4, P8, P12). Some participants found it simpler to use (P2, P5, P8) and helpful for catching up with the trainer (P12).
Adding the skeleton on top of the trainer (C4) was deemed as helpful for comparison (P1, P9), easier to adjust their pose (P11), and overall improved the performance (P10). It, however, made some parts of the trainer hard to see (P5, P7, P14) and was mentally challenging (P5) because of too much information (P10).


Participants reported that they looked mostly at the trainer for the whole experiment (P4-P7) or few first repetitions (P1-P3, P8-P12), limiting the time experience different conditions or comparison. Finally, participants have different learning styles, and each condition could be good in different situations and/or learning goals. For instance, skeleton could help getting a better idea of angle but make hands hidden (P5, P14). P3 would like to see the trainer video with skeleton first, then the user video with skeleton. P10 suggested that the tool should be goal oriented since little error is fine for self-practice so no need for detail comparison.



\section{Discussion and Limitations}
Our results suggest that C2 (trainer video + user video with skeleton) allowed our participants to be significantly more accurate. It was also deemed as less physically demanding and was overall preferred by our participants.
Data suggests that the user skeleton can help improving user performance. On its own, the user skeleton only offers limited benefits and makes comparison with the trainer data complicated, as shown by C3's lower performance. However, it allows the user to get a quick overview of their current pose.
The addition of the user's video feed also enabled a more precise side-by-side comparison with the trainer's pose. We were originally worried that this condition would be too demanding for the participants, but the NASA TLX results suggest that it was not the case, and that the benefit of making the tutorial video easier to follow offsets any potential additional cognitive load.
The skeleton on the trainer, on the other hand, could introduce visual clutter without improving user performance, as suggested by participants' comments and error of C3 versus C4. Visual clutter is not an issue in the user's case as the user already knows their own movement. Counter-intuitively, there is no need to visualize the trainer and user in the same way.


As a first step to explore automated processing of videos in the wild, we limited the case study to a video with few estimation errors. Our current system works well only when the trainer and user's whole body are viewed from the same camera angle. We also relied on an automatically estimated 2D skeleton, which may not be accurate in depicting 3D movement. We, however, believe that these limitations could be overcome in the near future by, for example, recent development in 3D human pose estimation from video~\cite{pavllo20193d,8949226}.

We found that some errors were based on a biased perception: for instance, P7 did not notice inaccuracy in their pose estimation until the experimenter pointed it out. They reasoned that since tai chi is slow, even one second of offset in terms of pose was unnoticeable when they focused on arm movement.


\section{Conclusion and Future Work}
In this paper, we investigated different types of visual feedback to help users follow tutorial videos. Our visual feedback relied on video feeds and pose estimation data. We found out that the best combination involves the trainer's video and the user's video feed superposed with pose estimation data. That type of feedback enabled direct comparison between both video feeds and allowed users to quickly assess their current pose using the skeleton. As future work, we would like to consider other applications such as dance and explore other mechanisms to help pose correction during training.
\bibliographystyle{ACM-Reference-Format}
\balance
\bibliography{references}


\end{document}